\documentclass[a4paper,11pt]{article}
\usepackage{jheppub}

\usepackage[utf8]{inputenc}
\usepackage{graphicx,cancel,float}
\usepackage{amssymb}
\usepackage{textcomp}
\usepackage{amsmath}
\usepackage{tabularx}
\usepackage{bm}
\usepackage{times}
\usepackage{color}
\usepackage{slashed}
\usepackage{multirow}
\usepackage{verbatim}
\usepackage{epsfig,epstopdf}
\usepackage{subfigure}

\allowdisplaybreaks[4]

\def\lsim{\mathrel{\raise.3ex\hbox{$<$\kern-.75em\lower1ex\hbox{$\sim$}}}}
\def\gsim{\mathrel{\raise.3ex\hbox{$>$\kern-.75em\lower1ex\hbox{$\sim$}}}}

\definecolor{orange}{rgb}{1,0.5,0}

\preprint{}
\subheader{\it \today}

\title{Solutions to axion electromagnetodynamics and new search strategies of sub-$\mu$eV axion}

\author[a]{Tong Li}
\emailAdd{litong@nankai.edu.cn}
\affiliation[a]{
School of Physics, Nankai University, Tianjin 300071, China
}
\author[a]{Rui-Jia Zhang}
\emailAdd{zhangruijia@mail.nankai.edu.cn}
\author[a]{Chang-Jie Dai}
\emailAdd{daichangjie@mail.nankai.edu.cn}

\abstract{
The Witten effect implies the electromagnetic interactions between axions and magnetic monopoles, and the quantum electromagnetodynamics (QEMD) properly describes interactions of electric charges, magnetic charges and photons. Based on the QEMD,
a generic low-energy axion-photon effective field theory was built by introducing two four-potentials ($A^\mu$ and $B^\mu$) to describe a photon. More anomalous axion-photon interactions and couplings ($g_{aAA}$, $g_{aBB}$ and $g_{aAB}$) arise in contrary to the ordinary axion coupling $g_{a\gamma\gamma}aF^{\mu\nu}\tilde{F}_{\mu\nu}$. As a consequence, the conventional axion Maxwell equations are further modified.
We properly solve the new axion-modified Maxwell equations and obtain the axion-induced electromagnetic fields given a static electric or magnetic field.
It turns out that the dominant couplings $g_{aAB}$ and $g_{aBB}$ can be probed in the presence of external magnetic field and electric field, respectively. The induced oscillating magnetic fields are always suppressed compared with the electric fields for the axions with large Compton wavelengths. This is contrary to the situation in conventional experiments searching for the oscillating magnetic fields induced by sub-$\mu$eV axions. Thus, we propose new strategies to measure the new couplings for sub-$\mu$eV axion in haloscope experiments.
}

\arxivnumber{arXiv:2211.06847}

\begin{document}

\maketitle
\setcounter{page}{2}

\newpage

\section{Introduction}
\label{sec:Intro}

Magnetic monopole and axion are two of the most interesting and mysterious candidates of physics beyond the Standard Model (SM).
Magnetic charges were initially motivated by the consideration of electric-magnetic symmetry in classical electromagnetism and Dirac suggested the existence of magnetic monopole in quantum theory in 1931~\cite{Dirac:1931kp}. The Dirac monopole was also generalized to those arising from QCD~\cite{Wu:1975es}, the grand unification theory~\cite{tHooft:1974kcl,Polyakov:1974ek} and the electroweak theory~\cite{Cho:1996qd}. Axions were introduced to solve the strong CP problem after the spontaneously breaking of Peccei-Quinn (PQ) symmetry~\cite{Baluni:1978rf,Crewther:1979pi,Kim:1979if,Shifman:1979if,Dine:1981rt,Zhitnitsky:1980tq,Baker:2006ts,Pendlebury:2015lrz} and have received a wide interest in both theoretical and experimental aspects. Both the QCD axion~\cite{Peccei:1977hh,Peccei:1977ur,Weinberg:1977ma,Wilczek:1977pj} (see Ref.~\cite{DiLuzio:2020wdo} for a recent review) and axion-like particles (ALPs)~\cite{Kim:1986ax,Kuster:2008zz} can play as dark matter (DM) through the misalignment mechanism~\cite{Preskill:1982cy,Dine:1982ah}.

In 1979, Witten pointed out that a non-zero vacuum angle $\theta$ in the CP violating term $\theta F^{\mu\nu}\tilde{F}_{\mu\nu}$ introduces an electric charge proportional to $\theta$ for magnetic monopoles~\cite{Witten:1979ey}. In axion theories, this Witten effect implies the electromagnetic interactions between axions and magnetic monopoles due to the axion-photon coupling $g_{a\gamma\gamma}a \vec{E}\cdot \vec{B}$. This connection was first derived by Fischler et al. under the semi-classical quantization of electromagnetism~\cite{Fischler:1983sc} and was proposed to solve various cosmological problems in recent years~\cite{Kawasaki:2015lpf,Nomura:2015xil,Kawasaki:2017xwt,Houston:2017kwe,Sato:2018nqy}. However, to properly quantize the axion-dyon dynamics in quantum field theory, one needs to utilize the quantum electromagnetodynamics (QEMD) built by Schwinger and Zwanziger~\cite{Schwinger:1966nj,Zwanziger:1968rs,Zwanziger:1970hk}. QEMD introduces two four-potentials ($A^\mu$ and $B^\mu$) and two $U(1)$ gauge groups ($U(1)_{\rm E}$ and $U(1)_{\rm M}$) to describe photons as well as electric and magnetic charges. Recently, based on quantization in QEMD, Ref.~\cite{Sokolov:2022fvs} constructed a generic axion-photon Lagrangian in the framework of low-energy axion effective field theory (EFT). It turns out that the interactions between axions and magnetic monopoles do exist in the absence of the Witten effect. More anomalous axion-photon interactions and couplings ($g_{aAA}$, $g_{aBB}$ and $g_{aAB}$) respecting shift symmetry arise in contrary to the ordinary axion EFT $g_{a\gamma\gamma} aF^{\mu\nu}\tilde{F}_{\mu\nu}$ in the SM framework. As a consequence of the above generic axion-photon Lagrangian, the classical equations of motion further modify the conventional axion Maxwell equations~\cite{Sikivie:1983ip}.

This framework predicts new phenomena induced by the new electromagnetic couplings of axions. Nowadays, various non-cavity haloscope experiments are proposed to search for
the ALPs with small masses $m_a\lesssim 1~\mu{\rm eV}$ and larger Compton wavelengths $\lambda_a$ than the physical scale of the detectors, such as ABRACADABRA~\cite{Kahn:2016aff,Salemi:2021gck}, ADMX SLIC~\cite{Crisosto:2019fcj}, DM Radio~\cite{Brouwer:2022bwo}, BASE~\cite{Devlin:2021fpq} and others. They search for the axion-induced oscillating magnetic field in the presence of a static magnetic field in a solenoid magnet~\cite{Sikivie:2013laa} or an external electric field~\cite{Gao:2020sjn,Gao:2022zxc,Bourhill:2022alm}, using an electronic LC circuit~\cite{Irastorza:2018dyq,Sikivie:2020zpn}. There also exist studies of the searches for axion-induced electric field~\cite{McAllister:2018ndu,Duan:2022nuy}. Nevertheless, to examine the detection of such low-mass ALPs, one needs to first solve the relevant axion Maxwell equations. In this work, inspired by the generic axion-photon couplings, we explore the solutions to QEMD-induced Maxwell equations and discuss the possibly new haloscope search strategies for the new axion couplings.

This paper is organized as follows. In Sec.~\ref{sec:QEMDandMaxwell}, we introduce the anomalous axion-photon interactions in QEMD and the modified Maxwell equations. In Sec.~\ref{sec:Solution}, we solve the Maxwell equations and give the axion induced electric and magnetic fields for experimental searches. The numerical results of dominant axion induced fields are shown in Sec.~\ref{sec:Exp}. Possible axion search experiments are also discussed. Our conclusions are drawn in Sec.~\ref{sec:Con}.

\section{The modified Maxwell equations from axion-photon interactions in QEMD}
\label{sec:QEMDandMaxwell}

\subsection{The anomalous axion-photon interactions in QEMD}
\label{sec:QEMD}

Ref.~\cite{Sokolov:2022fvs} builds the generic low-energy axion-photon EFT in the framework of QEMD. We briefly introduce the anomalous axion-photon interactions in QEMD below.

In the local QEMD, the photon is described by two four-potentials $A^\mu$ and $B^\mu$ with opposite parities. The $U(1)$ gauge group of QEMD correspondingly becomes $U(1)_{\rm E}\times U(1)_{\rm M}$ which inherently introduces both electric and magnetic charges.
The Lagrangian for the anomalous interactions between axion $a$ and photon in QEMD is~\cite{Sokolov:2022fvs}
\begin{eqnarray}
\mathcal{L}&\supset&  -{1\over 4} g_{aAA}~a~{\rm tr}[(\partial \wedge A)(\partial \wedge \tilde{A})] - {1\over 4} g_{aBB}~a~{\rm tr}[(\partial \wedge B)(\partial \wedge \tilde{B})]\nonumber \\
&& - {1\over 2} g_{aAB}~a~{\rm tr}[(\partial \wedge A)(\partial \wedge \tilde{B})]  \;,
\end{eqnarray}
where $(\partial \wedge X)^{\mu \nu}\equiv \partial^\mu X^\nu - \partial^\nu X^\mu$ for four-potential $X^\mu=A^\mu$ or $B^\mu$, and $(\partial \wedge \tilde{X})^{\mu \nu}\equiv \epsilon^{\mu\nu\rho\sigma} (\partial \wedge X)_{\rho\sigma}/2$ with $\epsilon^{0123}=-1$ as the Hodge dual tensor.
The first two dimension-five operators are CP-conserving axion interactions. Their couplings $g_{aAA}$ and $g_{aBB}$ are governed by the $U(1)_{\rm PQ}U(1)_{\rm E}^2$ and $U(1)_{\rm PQ}U(1)_{\rm M}^2$ anomalies, respectively. As $A^\mu$ and $B^\mu$ have opposite parities, the third operator is CP-violating one and its coupling $g_{aAB}$ is determined by the $U(1)_{\rm PQ}U(1)_{\rm E}U(1)_{\rm M}$ anomaly. It is analogous to the interaction between electromagnetic field and a scalar $\phi$ with positive parity $\phi F^{\mu\nu}F_{\mu\nu}$~\cite{Donohue:2021jbv}. The electromagnetic field strength tensors $F^{\mu\nu}$ and $\tilde{F}^{\mu\nu}$ are then introduced in the way that
\begin{eqnarray}
n\cdot F=n\cdot (\partial \wedge A)\;,~~ n\cdot \tilde{F}=n\cdot (\partial \wedge B)\;,
\end{eqnarray}
where $n^\mu = (0,\vec{n})$ is an arbitrary fixed spatial vector.

Taking care of the above anomalies, one can calculate the coupling coefficients as
\begin{eqnarray}
g_{aAA}={Ee^2\over 4\pi^2 v_{\rm PQ}}\;,~~g_{aBB}={Mg_0^2\over 4\pi^2 v_{\rm PQ}}\;,~~g_{aAB}={Deg_0\over 4\pi^2 v_{\rm PQ}}\;,
\label{eq:couplings}
\end{eqnarray}
where $e$ is the unit of electric charge, $g_0$ is the minimal magnetic charge with $g_0=2\pi/e$ in the Dirac-Schwinger-Zwanziger (DSZ) quantization condition, and $v_{\rm PQ}$ is the $U(1)_{\rm PQ}$ symmetry breaking scale. $E(M)$ is the electric (magnetic) anomaly coefficient and $D$ is the mixed electric-magnetic CP-violating anomaly coefficient. They are computed by integrating out heavy PQ-charged fermions with electric and magnetic charges. As the DSZ quantization condition indicates $g_0\gg e$, we have the scaling of the axion-photon couplings as $g_{aBB}\gg |g_{aAB}|\gg g_{aAA}$.
We summarize the details of QEMD and the calculation of the anomaly coefficients in Appendix.

\subsection{The modified Maxwell equations}
\label{sec:Maxwell}

Given the above axion-photon interactions as well as the free Lagrangian, one can derive the classical equations of motion.
The conventional axion-electrodynamics is then modified.
The axion modified Maxwell equations are newly obtained as~\cite{Sokolov:2022fvs}
\begin{eqnarray}
&&\vec{\nabla}\times \vec{B}_a-{\partial \vec{E}_a\over \partial t}=g_{aAA}(\vec{E}_0 \times \vec{\nabla} a - {\partial a\over \partial t} \vec{B}_0)
+ g_{aAB} (\vec{B}_0 \times \vec{\nabla} a + {\partial a\over \partial t} \vec{E}_0)\;,\\
&&\vec{\nabla}\times \vec{E}_a+{\partial \vec{B}_a\over \partial t}=-g_{aBB}(\vec{B}_0 \times \vec{\nabla} a + {\partial a\over \partial t} \vec{E}_0)
- g_{aAB} (\vec{E}_0 \times \vec{\nabla} a - {\partial a\over \partial t} \vec{B}_0)\;,\\
&&\vec{\nabla}\cdot \vec{B}_a = -g_{aBB} \vec{E}_0\cdot \vec{\nabla} a + g_{aAB} \vec{B}_0\cdot \vec{\nabla} a \;,\\
&&\vec{\nabla}\cdot \vec{E}_a = g_{aAA} \vec{B}_0\cdot \vec{\nabla} a - g_{aAB} \vec{E}_0\cdot \vec{\nabla} a \;,
\end{eqnarray}
and the new Klein-Gordon equation is
\begin{eqnarray}
(\Box+m_a^2)a=(g_{aAA}+g_{aBB})\vec{E}_0\cdot \vec{B}_0 + g_{aAB}(\vec{E}_0^2-\vec{B}_0^2)\;,
\end{eqnarray}
where $\vec{E}_0$ and $\vec{B}_0$ are static electric and magnetic fields in a detector, and $\vec{E}_a$ and $\vec{B}_a$ are axion-induced electric and magnetic fields. Note that one has expanded the electromagnetic field up to the first order of axion-photon couplings and omitted the parts of ordinary Maxwell equations in the above equations.
When taking $g_{aBB}=g_{aAB}=0$ and replacing $g_{aAA}$ by the conventional coupling $g_{a\gamma\gamma}$, the above equations restore to the conventional axion modified Maxwell equations~\cite{Sikivie:1983ip}.

Based on Eq.~(\ref{eq:couplings}), assuming the coefficients $E\simeq M\simeq |D|$, we find $g_{aAA}/g_{aBB}\simeq (e/g_0)^2 \simeq 10^{-4}$ and $|g_{aAB}|/g_{aBB}\simeq e/g_0 \simeq 10^{-2}$. Also, the axion dark matter has a typical local velocity $v_{\rm DM}=|\vec{v}_a|\sim 10^{-3}c$ in the Milky Way and then one has $|\vec{\nabla} a|\sim 10^{-3} \partial a/\partial t$. As a result, keeping only the first three dominant terms simplifies the above Maxwell equations.
The simplified Maxwell equations become
\begin{eqnarray}
&&\vec{\nabla}\times \vec{B}_a-{\partial \vec{E}_a\over \partial t}=0\;,\label{eq:ecu}\\
&&\vec{\nabla}\times \vec{E}_a+{\partial \vec{B}_a\over \partial t}=-g_{aBB}(\vec{B}_0 \times \vec{\nabla} a + {\partial a\over \partial t} \vec{E}_0)
+ g_{aAB} {\partial a\over \partial t} \vec{B}_0\;,\label{eq:mcu}\\
&&\vec{\nabla}\cdot \vec{B}_a = 0 \;,\label{eq:mch}\\
&&\vec{\nabla}\cdot \vec{E}_a = 0 \;. \label{eq:cch}
\end{eqnarray}
These are the wave equations that we will solve in next section.

\section{Solutions to axion electromagnetodynamics}
\label{sec:Solution}

\subsection{Case I: $\vec{B}_0\neq 0$ and $\vec{E}_0=0$}
\label{sec:caseI}

\begin{figure}[htb!]
\centering
\includegraphics[scale=0.7]{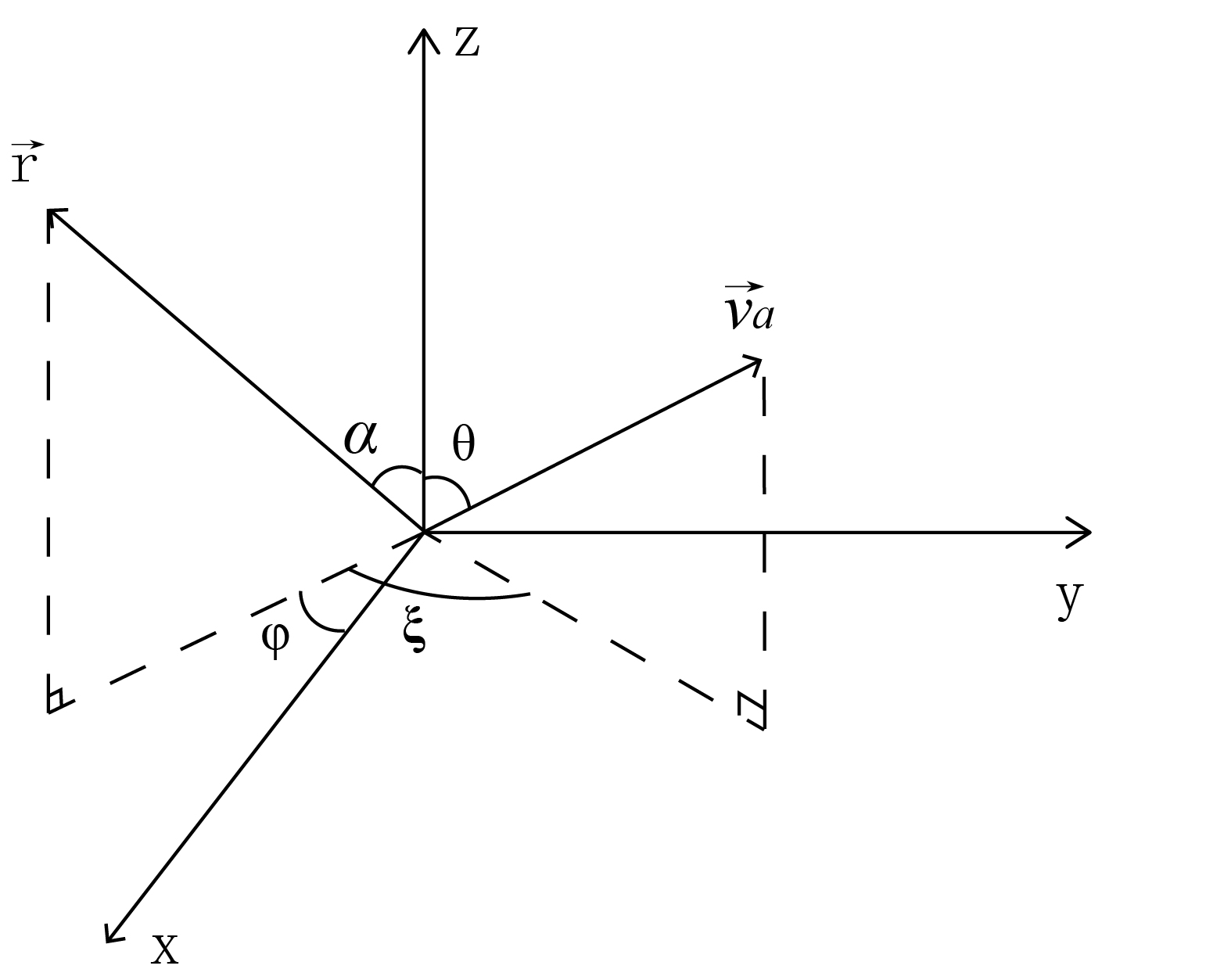}
\caption{The coordinates of axion $\vec{v}_a$ and $\vec{r}$.}
\label{fig:coordinates}
\end{figure}

The ordinary haloscope experiments adopt an external magnetic field $\vec{B}_0\neq 0$ but vanishing electric field $\vec{E}_0=0$.
In contrary to the conventional axion modified Maxwell equations, Eq.~(\ref{eq:mcu}) induces an effective magnetic current: $\vec{j}^m_{\rm eff}=g_{aBB}\vec{B}_0 \times \vec{\nabla} a - g_{aAB} {\partial a\over \partial t} \vec{B}_0$. After applying the curl differential operator to the Eqs.~(\ref{eq:ecu}) and (\ref{eq:mcu}), in the case with $\vec{B}_0\neq 0$ and $\vec{E}_0 = 0$, one can obtain
\begin{eqnarray}
\nabla^2 \vec{B}_a - {\partial^2 \vec{B}_a\over \partial t^2}&=&g_{aBB}\vec{B}_0\times \vec{\nabla} {\partial a\over \partial t} - g_{aAB}{\partial^2 a\over \partial t^2} \vec{B}_0\;, \label{eq:Beq}\\
\nabla^2 \vec{E}_a - {\partial^2 \vec{E}_a\over \partial t^2}&=&g_{aBB} (\vec{\nabla}a\cdot \vec{\nabla})\vec{B}_0 - g_{aAB}{\partial a\over \partial t} \vec{\nabla}\times \vec{B}_0\;. \label{eq:Eeq}
\end{eqnarray}
To solve Eqs.~(\ref{eq:Beq}) and (\ref{eq:Eeq}), we take a simple geometry of a long solenoid with a radius $R$ and a static magnetic field along the $z$ direction in cylindrical coordinates $(\rho, \phi, z)$. The magnetic field around the solenoid is parameterized as $\vec{B}_0=\theta(R-\rho)B_0 \hat{z}$ with the Heaviside theta function $\theta(x)$.
Then, Eq.~(\ref{eq:Eeq}) becomes
\begin{eqnarray}
\nabla^2 \vec{E}_a - {\partial^2 \vec{E}_a\over \partial t^2}&=& - \Big(g_{aBB}{\partial a\over \partial \rho}\hat{z}+g_{aAB}{\partial a\over \partial t}\hat{\phi}\Big)B_0 \delta(\rho-R)\;.
\label{eq:Eeqre}
\end{eqnarray}
The axion field is given by $a(t,\vec{r})=a_0\cos(\omega_a t - \vec{k}_a\cdot \vec{r})$ with $\omega_a=m_a$ and $\vec{k}_a=m_a \vec{v}_a$.
We parameterize the direction of axion in spherical coordinates with the angles shown in Fig.~\ref{fig:coordinates} and then we have $\vec{v}_a=v_a(\sin\theta\cos(\xi-\phi),\sin\theta\sin(\xi-\phi),\cos\theta)$.

Now we follow Ref.~\cite{Ouellet:2018nfr}~\footnote{Another calculation based on quantum field theory was given in Ref.~\cite{Beutter:2018xfx}.} to solve Eq.~(\ref{eq:Eeqre}) in $\phi$ direction and propose the solution as $\vec{E}_a=U_{E\phi}(\rho)e^{i\omega_a t}\hat{\phi}$.
After inserting this solution form into the $\hat{\phi}$ component of Eq.~(\ref{eq:Eeqre}), we obtain the following Bessel equation
\begin{eqnarray}
\Big[ \partial^2_{\rho'}+{1\over \rho'}\partial_{\rho'}+\Big(1-{1\over \rho^{\prime 2}}\Big) \Big]U_{E\phi}(\rho')=-i g_{aAB} a_0 B_0 \delta(\rho'-\omega_a R)\;,
\end{eqnarray}
where $\rho'=\omega_a \rho$.
With the boundary conditions at $\rho'=0$ and $\rho'=\omega_a R$, the solutions to the above equation are Bessel functions of order one
\begin{eqnarray}
U_{E\phi}(\rho')=\left\{
             \begin{array}{ll}
               a_{E\phi} J_1(\rho'), & \rho'<\omega_a R\; , \\
               b_{E\phi} H_1^+(\rho'), & \rho'>\omega_a R\; ,
             \end{array}
           \right.
\end{eqnarray}
where $J_1(\rho')$ is the spherical Bessel function of the first kind and $H_1^+(\rho')$ is the spherical Hankel function of the first kind describing outgoing wave.
Utilizing the continuity of electric field $U_{E\phi}(\rho')$ and the discontinuity of $\partial U_{E\phi}/\partial \rho'$ across the boundary, we obtain the equations for the coefficients $a_E$ and $b_E$
\begin{eqnarray}
&&a_{E\phi} J_1(\omega_a R) - b_{E\phi} H_1^+ (\omega_a R) =0\;,\\
&&\Big[b_{E\phi} {\partial H_1^+\over \partial \rho'} - a_{E\phi} {\partial J_1\over \partial \rho'} \Big]_{\rho'=\omega_a R} =  -i g_{aAB} a_0 B_0\;.
\end{eqnarray}
After applying the Wronksian of Bessel functions, the coefficients are obtained as
\begin{eqnarray}
a_{E\phi} &=& -{\pi\over 2} g_{aAB} a_0 B_0 \omega_a R H_1^+ (\omega_a R)\;,\\
b_{E\phi} &=& -{\pi\over 2} g_{aAB} a_0 B_0 \omega_a R J_1 (\omega_a R) \;.
\end{eqnarray}
Considering the limit of large Compton wavelengths $\lambda_a\gg R$ and thus $\rho'=\omega_a R\ll 1$, the above Bessel functions can be simplified. The final solutions of $\vec{E}_a$ in $\phi$ direction become
\begin{eqnarray}
\vec{E}_{a,\phi}&\approx &\left\{
            \begin{array}{ll}
              i\Big[{1\over 2}g_{aAB}a_0 B_0 \omega_a \rho - {1\over 4}g_{aAB}a_0 B_0 \omega_a^3 R^2\rho \Big(\gamma'(\omega_a R)-{1\over 2}\Big) \Big]e^{i\omega_a t} \hat{\phi}, & \rho<R \;,\\
              i\Big[{1\over 2}g_{aAB}a_0 B_0 \omega_a {R^2\over \rho} - {1\over 4}g_{aAB}a_0 B_0 \omega_a^3 R^2\rho \Big(\gamma'(\omega_a R)-{1\over 2}\Big) \Big]e^{i\omega_a t} \hat{\phi}, & \rho>R \;,
            \end{array}
          \right. \\
          &\approx &\left\{
            \begin{array}{ll}
              i{1\over 2}g_{aAB}a_0 B_0 \omega_a \rho e^{i\omega_a t} \hat{\phi}, & \rho<R \;,\\
              i{1\over 2}g_{aAB}a_0 B_0 \omega_a {R^2\over \rho} e^{i\omega_a t} \hat{\phi}, & \rho>R \;,
            \end{array}
          \right.
\end{eqnarray}
where $\gamma'(x)={\rm ln}(x/2)+\gamma-i\pi/2$ with the Euler-Mascheroni constant being $\gamma\approx 0.5772$.

Then we take $\vec{E}_a=U_{Ez}(\rho)e^{i\omega_a t}\hat{z}$ and follow the same procedure to solve the electric field in $z$ direction.
The corresponding Bessel equation of order zero is
\begin{eqnarray}
\Big[ \partial^2_{\rho'}+{1\over \rho'}\partial_{\rho'}+1 \Big]U_{Ez}(\rho')=i g_{aBB} a_0 B_0 v_a \sin\theta \cos(2\phi-\xi) \delta(\rho'-\omega_a R)\;.
\end{eqnarray}
The solutions are given by
\begin{eqnarray}
U_{Ez}(\rho')=\left\{
             \begin{array}{ll}
               a_{Ez} J_0(\rho'), & \rho'<\omega_a R\; , \\
               b_{Ez} H_0^+(\rho'), & \rho'>\omega_a R\; .
             \end{array}
           \right.
\end{eqnarray}
Given the boundary conditions, the coefficients satisfy
\begin{eqnarray}
&&a_{Ez} J_0(\omega_a R) - b_{Ez} H_0^+ (\omega_a R) =0\;,\\
&&\Big[b_{Ez} {\partial H_0^+\over \partial \rho'} - a_{Ez} {\partial J_0\over \partial \rho'} \Big]_{\rho'=\omega_a R} =  i g_{aBB} a_0 B_0 v_a \sin\theta \cos(2\phi-\xi)\;,
\end{eqnarray}
and the solutions become
\begin{eqnarray}
a_{Ez}&=&{\pi\over 2}g_{aBB}a_0 B_0 v_a \sin\theta \cos(2\phi-\xi) \omega_a R H_0^+(\omega_a R)\;,\\
b_{Ez}&=&{\pi\over 2}g_{aBB}a_0 B_0 v_a \sin\theta \cos(2\phi-\xi) \omega_a R J_0(\omega_a R)\;.
\end{eqnarray}
The final solutions of $\vec{E}_a$ in $z$ direction are
\begin{eqnarray}
\vec{E}_{a,z}&\approx &\left\{
            \begin{array}{ll}
              ig_{aBB}a_0 B_0 v_a\omega_a R \Big[\gamma'(\omega_a R)\Big(1-{\omega_a^2 \rho^2\over 4}\Big)\\
              +{1\over 4}(1-\gamma'(\omega_a R))(\omega_a R)^2\Big]\sin\theta\cos(2\phi-\xi) e^{i\omega_a t} \hat{z}, & \rho<R \;,\\
              ig_{aBB}a_0 B_0 v_a\omega_a R \Big[\gamma'(\omega_a \rho)\Big(1-{\omega_a^2 R^2\over 4}\Big)\\
              +{1\over 4}(1-\gamma'(\omega_a \rho))(\omega_a \rho)^2\Big]\sin\theta\cos(2\phi-\xi) e^{i\omega_a t} \hat{z}, & \rho>R \;.
            \end{array}
          \right. 
\end{eqnarray}

Next we solve the magnetic field $\vec{B}_a$. As the first term on the right-handed side of Eq.~(\ref{eq:Beq}) is perpendicular to $\vec{B}_0$,
only the second term contributes to the wave equation in $z$ direction as
\begin{eqnarray}
\nabla^2 \vec{B}_a - {\partial^2 \vec{B}_a\over \partial t^2} = -g_{aAB} B_0 {\partial^2 a\over \partial t^2}\theta(R-\rho) \hat{z} \;.
\end{eqnarray}
We propose the solution as $\vec{B}_a=U_{Bz}(\rho)e^{i\omega_a t}\hat{z}$ and the Bessel equation is then
\begin{eqnarray}
\Big[ \partial^2_{\rho'}+{1\over \rho'}\partial_{\rho'}+1 \Big]U_{Bz}(\rho')=g_{aAB}a_0 B_0 \theta(\omega_a R-\rho')\;.
\end{eqnarray}
The solutions are
\begin{eqnarray}
U_{Bz}(\rho')=\left\{
             \begin{array}{ll}
               a_{Bz} J_0(\rho') + g_{aAB}a_0 B_0, & \rho'<\omega_a R\; , \\
               b_{Bz} H_0^+(\rho'), & \rho'>\omega_a R\; ,
             \end{array}
           \right.
\end{eqnarray}
with the coefficients as
\begin{eqnarray}
a_{Bz}&=&-{i\pi\over 2} g_{aAB} a_0 B_0 \omega_a R H_1^+(\omega_a R)\;,\\
b_{Bz}&=&-{i\pi\over 2} g_{aAB} a_0 B_0 \omega_a R J_1(\omega_a R)\;.
\end{eqnarray}
We find the $\vec{B}_a$ solutions in $z$ direction are
\begin{eqnarray}
\vec{B}_{a,z}&\approx &\left\{
            \begin{array}{ll}
              g_{aAB}a_0 B_0\Big[{(\omega_a R)^2\over 2}\Big(\gamma'(\omega_a R)-{1\over 2}\Big)\Big(1-{\omega_a^2\rho^2\over 4}\Big)+{\omega_a^2\rho^2\over 4}\Big]e^{i\omega_a t} \hat{z}, & \rho<R \;,\\
              g_{aAB}a_0 B_0 {(\omega_a R)^2\over 2}\Big[\gamma'(\omega_a \rho)+ {1\over 4}(1-\gamma'(\omega_a \rho))(\omega_a \rho)^2\Big]e^{i\omega_a t} \hat{z}, & \rho>R \;,
            \end{array}
          \right. \\
          &\approx &\left\{
            \begin{array}{ll}
              g_{aAB}a_0 B_0\Big[{(\omega_a R)^2\over 2}\Big(\gamma'(\omega_a R)-{1\over 2}\Big)+{\omega_a^2\rho^2\over 4}\Big]e^{i\omega_a t} \hat{z}, & \rho<R \;,\\
              g_{aAB}a_0 B_0 {(\omega_a R)^2\over 2}\gamma'(\omega_a \rho) e^{i\omega_a t} \hat{z}, & \rho>R \;.
            \end{array}
          \right.
\end{eqnarray}

For the $\vec{B}_a$ field in $\phi$ direction, we have the equation as
\begin{eqnarray}
\nabla^2 \vec{B}_a - {\partial^2 \vec{B}_a\over \partial t^2} = g_{aBB} B_0 v_a \omega_a^2 a \sin\theta \cos(2\phi-\xi)\theta(R-\rho) \hat{\phi} \;.
\end{eqnarray}
Inserting the solution $\vec{B}_a=U_{B\phi}(\rho)e^{i\omega_a t}\hat{\phi}$, the Bessel equation of order one becomes
\begin{eqnarray}
\Big[ \partial^2_{\rho'}+{1\over \rho'}\partial_{\rho'}+\Big(1-{1\over \rho^{\prime 2}}\Big) \Big]U_{B\phi}(\rho')= g_{aBB} a_0 B_0 v_a \sin\theta \cos(2\phi-\xi) \theta(\omega_a R-\rho')\;.
\end{eqnarray}
It turns out to be a nonhomogeneous Bessel equation of order one when $\rho<R$.
We use the software Mathematica to find the solutions as
\begin{eqnarray}
U_{B\phi}(\rho')=\left\{
             \begin{array}{ll}
               a_{B\phi}J_1(\rho')+{k\pi\over 12}\rho^{\prime 3}Y_1(\rho')H(\rho')-k\pi J_1(\rho')M(\rho'), & \rho'<\omega_a R\; , \\
               b_{B\phi} H_1^+(\rho'), & \rho'>\omega_a R\; ,
             \end{array}
           \right.
\end{eqnarray}
where $k=g_{aBB} a_0 B_0 v_a \sin\theta \cos(2\phi-\xi)$, $H(x)$ is the generalized hypergeometric function and $M(x)$ is the Meijer G function
\begin{eqnarray}
H(x)&=&{\rm HypergeometricPFQ}[\{{3\over 2}\},\{2,{5\over 2}\},-{x^2\over 4}]\;,\\
M(x)&=&{\rm MeijerG}[\{\{1\},\{0\}\},\{\{{1\over 2},{3\over 2}\} , \{0,0\}\},{x\over 2},{1\over 2}]\;.
\end{eqnarray}
Using the boundary conditions, the coefficients are given by
\begin{eqnarray}
a_{B\phi}&=&-{i\over 12}k\pi(\omega_a R)^3 H(\omega_a R)+k\pi M(\omega_a R)\;,\\
b_{B\phi}&=&-{i\over 12}k\pi(\omega_a R)^3 H(\omega_a R)\;.
\end{eqnarray}
Then, the $\vec{B}_a$ solutions in $\phi$ direction are
\begin{eqnarray}
\vec{B}_{a,\phi}&\approx &\left\{
            \begin{array}{ll}
              k\Big[-{i\over 24}\pi \omega_a^4 R^3\rho H(\omega_a R)+{\pi\over 2}\omega_a\rho M(\omega_a R)+{1\over 12}\omega_a^4\rho^4 ({\rm ln}(\omega_a\rho)+\gamma-{1\over 2})H(\omega_a\rho)\\
              -{1\over 6}\omega_a^2\rho^2H(\omega_a\rho)-{\pi\over 2}\omega_a\rho M(\omega_a \rho)\Big]e^{i\omega_a t} \hat{\phi}, & \rho<R \;,\\
              {1\over 12} k \Big[\omega_a^4 R^3\rho (\gamma'(\omega_a\rho)-{1\over 2})H(\omega_a R)-2\omega_a^2{R^3\over \rho}H(\omega R)\Big] e^{i\omega_a t} \hat{\phi}, & \rho>R \;.
            \end{array}
          \right.
\label{eq:Baphi}
\end{eqnarray}

The equation of the $\vec{B}_a$ field in $\rho$ direction is
\begin{eqnarray}
\nabla^2 \vec{B}_a - {\partial^2 \vec{B}_a\over \partial t^2} = 2g_{aBB} B_0 v_a \omega_a^2 a \sin\theta \sin(2\phi-\xi)\theta(R-\rho) \hat{\rho} \;.
\end{eqnarray}
One can see that it is analogous to the equation for $\vec{B}_{a,\phi}$. To obtain solutions of the magnetic field, we only need to replace the value of $k$ in Eq.~(\ref{eq:Baphi}) by $k=2g_{aBB}a_0 B_0 v_a \sin\theta\sin(2\phi-\xi)$.

The dominant axion electromagnetic fields here are $\vec{E}_{a,\phi}$ and $\vec{B}_{a,z}$ without velocity $v_a$ suppression. They are equivalent to the solutions of conventional axion-modified Maxwell equations in Ref.~\cite{Ouellet:2018nfr} by replacing $\vec{E}_{a,\phi}\to \vec{B}_{a,\phi}$, $\vec{B}_{a,z}\to -\vec{E}_{a,z}$ and $g_{aAB}\to g_{a\gamma\gamma}$ in our results.

\subsection{Case II: $\vec{B}_0 = 0$ and $\vec{E}_0\neq 0$}
\label{sec:caseII}

In the case with $\vec{B}_0= 0$ and $\vec{E}_0 \neq 0$, the wave equations become
\begin{eqnarray}
\nabla^2 \vec{B}_a - {\partial^2 \vec{B}_a\over \partial t^2}&=& g_{aBB}{\partial^2 a\over \partial t^2} \vec{E}_0\;, \label{eq:Beq1}\\
\nabla^2 \vec{E}_a - {\partial^2 \vec{E}_a\over \partial t^2}&=& g_{aBB}{\partial a\over \partial t} \vec{\nabla}\times \vec{E}_0\;. \label{eq:Eeq1}
\end{eqnarray}
They can be rewritten as
\begin{eqnarray}
\nabla^2 \vec{B}_a - {\partial^2 \vec{B}_a\over \partial t^2} &=& g_{aBB} E_0 {\partial^2 a\over \partial t^2}\theta(R-\rho) \hat{z} \;,\\
\nabla^2 \vec{E}_a - {\partial^2 \vec{E}_a\over \partial t^2}&=& g_{aBB} E_0 {\partial a\over \partial t}\delta(\rho-R)\hat{\phi}\;.
\end{eqnarray}
Following the same procedures in the above subsection, we obtain the dominant $\vec{E}_a$ in $\phi$ direction as
\begin{eqnarray}
\vec{E}_{a,\phi}&=&\left\{
            \begin{array}{ll}
              {\pi\over 2}g_{aBB}a_0 E_0 \omega_a R H_1^+(\omega_a R) J_1(\omega_a \rho) e^{i\omega_a t} \hat{\phi}, & \rho<R \;,\\
              {\pi\over 2}g_{aBB}a_0 E_0 \omega_a R J_1(\omega_a R) H_1^+(\omega_a \rho)  e^{i\omega_a t} \hat{\phi}, & \rho>R \;,
            \end{array}
          \right.\\
          &\approx &\left\{
            \begin{array}{ll}
              -i{1\over 2}g_{aBB}a_0 E_0 \omega_a \rho e^{i\omega_a t} \hat{\phi}, & \rho<R \;,\\
              -i{1\over 2}g_{aBB}a_0 E_0 \omega_a {R^2\over \rho} e^{i\omega_a t} \hat{\phi}, & \rho>R \;.
            \end{array}
          \right.
\end{eqnarray}
The solution of dominant $\vec{B}_a$ in $z$ direction is
\begin{eqnarray}
\vec{B}_{a,z}&=&\left\{
            \begin{array}{ll}
              \Big[i{\pi\over 2}g_{aBB}a_0 E_0 \omega_a R H_1^+(\omega_a R) J_0(\omega_a \rho)-g_{aBB}a_0 E_0\Big] e^{i\omega_a t} \hat{z}, & \rho<R \;,\\
              i{\pi\over 2}g_{aBB}a_0 E_0 \omega_a R J_1(\omega_a R) H_0^+(\omega_a \rho) e^{i\omega_a t} \hat{z}, & \rho>R \;,
            \end{array}
          \right.\\
          &\approx &\left\{
            \begin{array}{ll}
              -g_{aBB}a_0 E_0\Big[{(\omega_a R)^2\over 2}\Big(\gamma'(\omega_a R)-{1\over 2}\Big)+{\omega_a^2\rho^2\over 4}\Big]e^{i\omega_a t} \hat{z}, & \rho<R \;,\\
              -g_{aBB}a_0 E_0 {(\omega_a R)^2\over 2}\gamma'(\omega_a \rho) e^{i\omega_a t} \hat{z}, & \rho>R \;.
            \end{array}
          \right.
\end{eqnarray}

\section{Numerical results and new haloscope experiments}
\label{sec:Exp}

Based on the above analytical results, one finds that the dominant couplings $g_{aAB}$ and $g_{aBB}$ can be probed in the presence of external magnetic field and electric field, respectively. Moreover, the induced oscillating magnetic fields are suppressed compared with the electric fields for the axions with large Compton wavelengths $\lambda_a=2\pi/m_a\gg R$. The electric field $\vec{E}_a$ in $\phi$ direction is always dominant. This is contrary to the situation in conventional experiments searching for the oscillating magnetic fields induced by sub-$\mu$eV axions.
In this section, we show the numerical results to demonstrate the size of induced electromagnetic fields and propose new strategies to measure the oscillating electric fields.

\subsection{Numerical results of axion-induced electromagnetic fields}

\begin{figure}[htb!]
\centering
\includegraphics[scale=0.8]{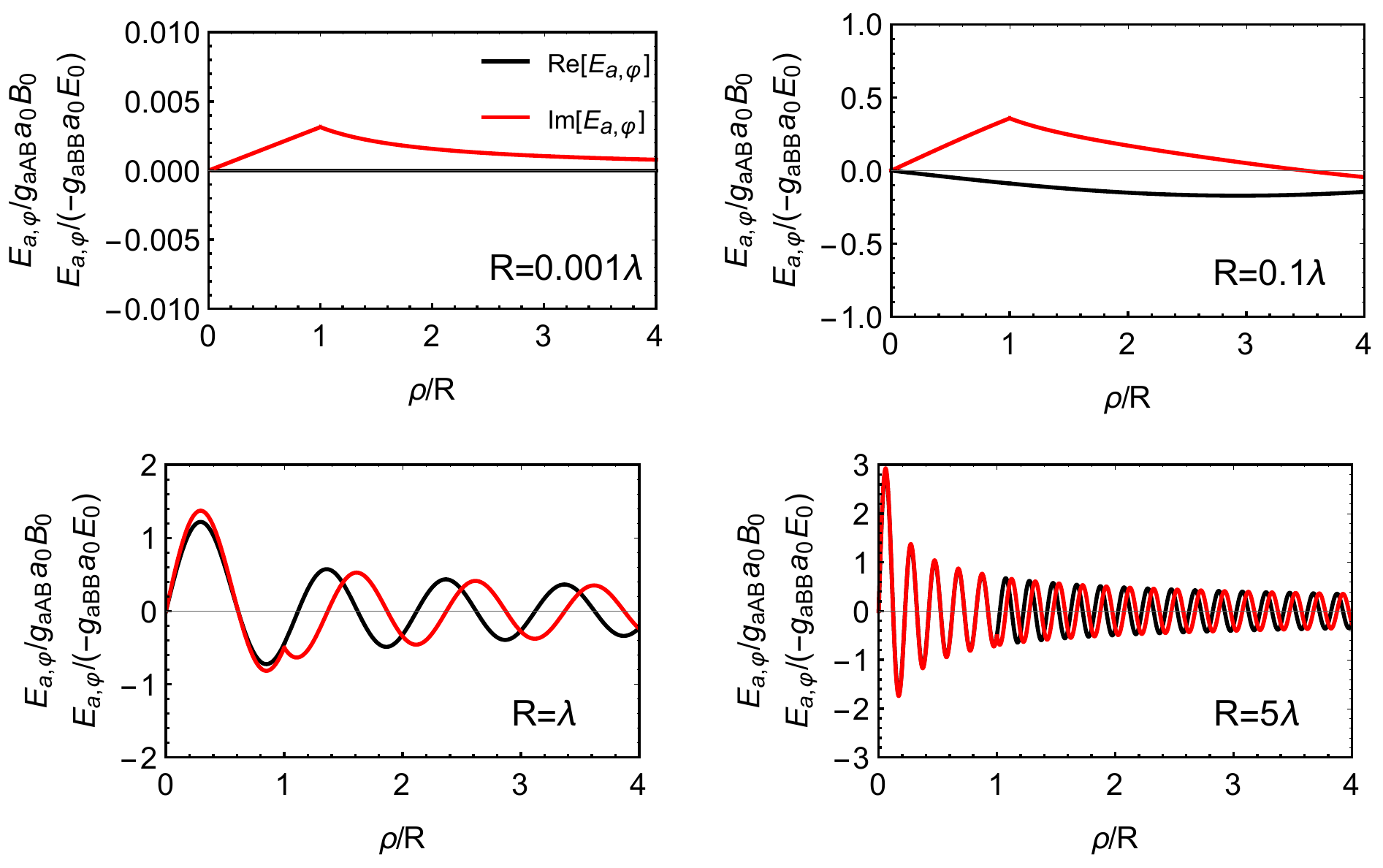}
\caption{Numerical results of axion-induced oscillating electric field $\vec{E}_{a,\phi}$ at $t=0$, in units of $g_{aAB}a_0B_0$ for case I or $-g_{aBB}a_0E_0$ for case II. We consider four kinds of relations between the detector scale $R$ and axion Compton wavelength $\lambda_a$: $R = 0.001\lambda_a$ (top left),
$0.1\lambda_a$ (top right), $\lambda_a$ (bottom left) and $5\lambda_a$ (bottom right), where $\lambda_a=2\pi/m_a$.
}
\label{fig:E_phi}
\end{figure}

\begin{figure}[htb!]
\centering
\includegraphics[scale=0.8]{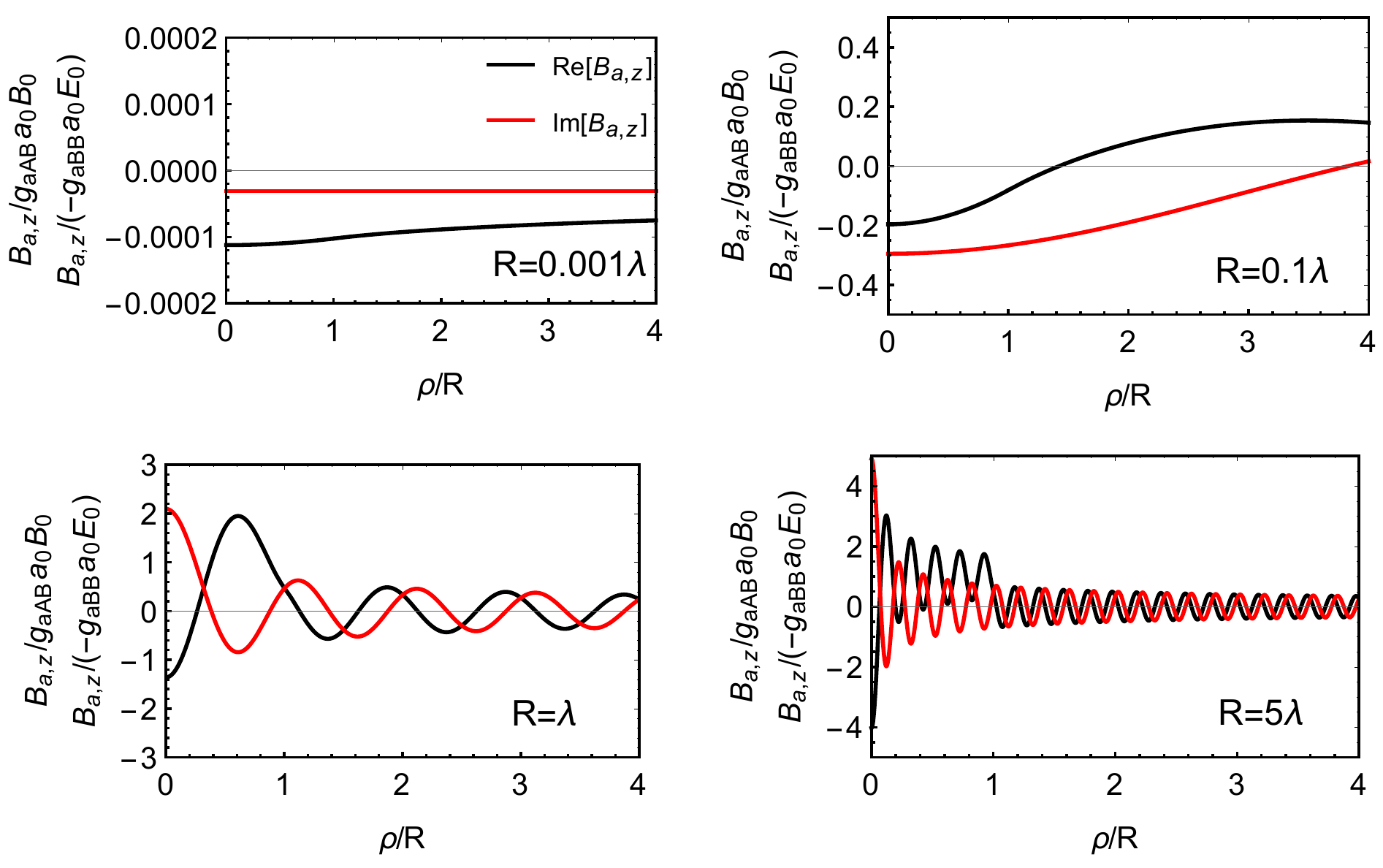}
\caption{Numerical results of axion-induced oscillating magnetic field $\vec{B}_{a,z}$ at $t=0$, in units of $g_{aAB}a_0B_0$ for case I or $-g_{aBB}a_0E_0$ for case II, as labeled in Fig.~\ref{fig:E_phi}.
}
\label{fig:B_z}
\end{figure}

In the case I with $\vec{B}_0\neq 0$ and $\vec{E}_0= 0$, as shown in Sec.~\ref{sec:caseI}, the axion-induced electromagnetic fields proportion to $g_{aBB}$ are suppressed by the velocity of axion DM $v_a\sim10^{-3}$. It is clear that the components $\vec{E}_{a,\phi}$ and $\vec{B}_{a,z}$ determined by coupling $g_{aAB}$ are dominant.
We numerically evaluate the results in Sec.~\ref{sec:caseI}. The distributions of field strength $\vec{E}_{a,\phi}$ and $\vec{B}_{a,z}$ as a function of ratio $\rho/R$ are displayed in Fig.~\ref{fig:E_phi} and Fig.~\ref{fig:B_z}, respectively.

In the limit of $\lambda_a\gg R$, we find that $E_{a,\phi}$ is about one order of magnitude larger than $B_{a,z}$ under the long wavelength approximation ($R=0.001\lambda_a$). While in other cases with much lower wavelengths ($R=0.1, 1$ and $5 \lambda_a$), the electromagnetic fields begin to oscillate due to the Bessel function in the field solutions and they have no significant difference. Consequently, contrary to the usual method searching for axion-induced oscillating magnetic field $B_a$ in $z$ direction in the present axion haloscope experiments, it is a reasonable way to measure the coupling $g_{aAB}$ by searching for the induced electric field $E_{a,\phi}$ via an external magnetic field $B_0$.

To measure the coupling $g_{aBB}$, as discussed in Sec.~\ref{sec:caseII}, we consider a uniform electric field $E_0$ along z-axis and spatially parameterized by $\rho$. In this case, the field solutions are analogous to the results of $E_{a,\phi}$ and $B_{a,z}$ in case I, only differing by the substitution of $g_{aAB}B_0 \to -g_{aBB}E_0$ as shown in Figs.~\ref{fig:E_phi} and \ref{fig:B_z}. Thus, in this case, searching for the induced electric field is still a proper approach to probe the signal of axion field even in the external electric field $E_0$.

\subsection{New search strategies of sub-$\mu$eV axion}

\begin{figure}[htb!]
\centering
\includegraphics[scale=0.15]{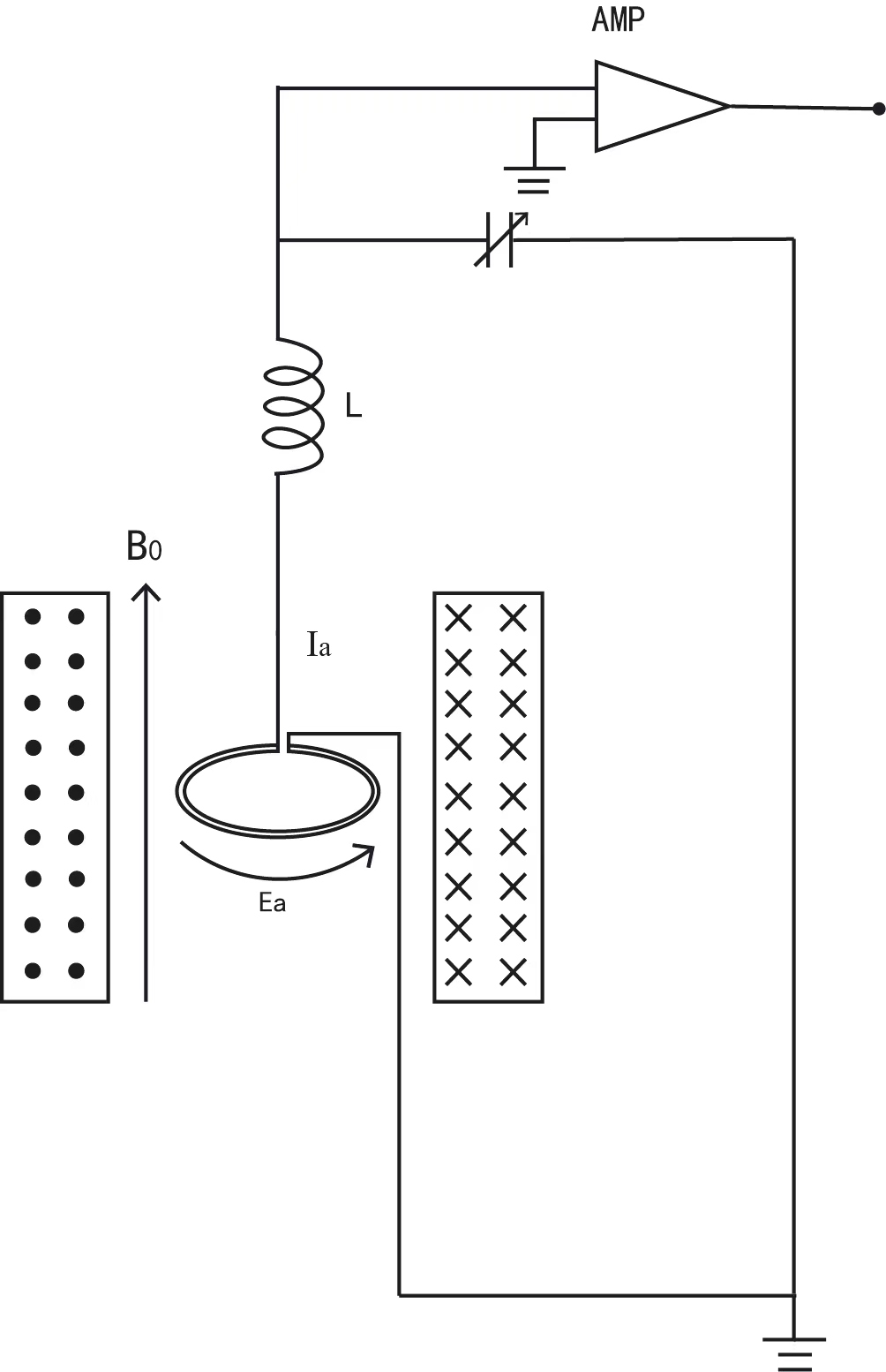}
\caption{The schematic diagram of experimental setup for case I. For case II, the external solenoid is replaced by horizontally placed parallel plates.
}
\label{fig:setup}
\end{figure}

\begin{figure}[htb!]
\centering
\includegraphics[scale=0.6]{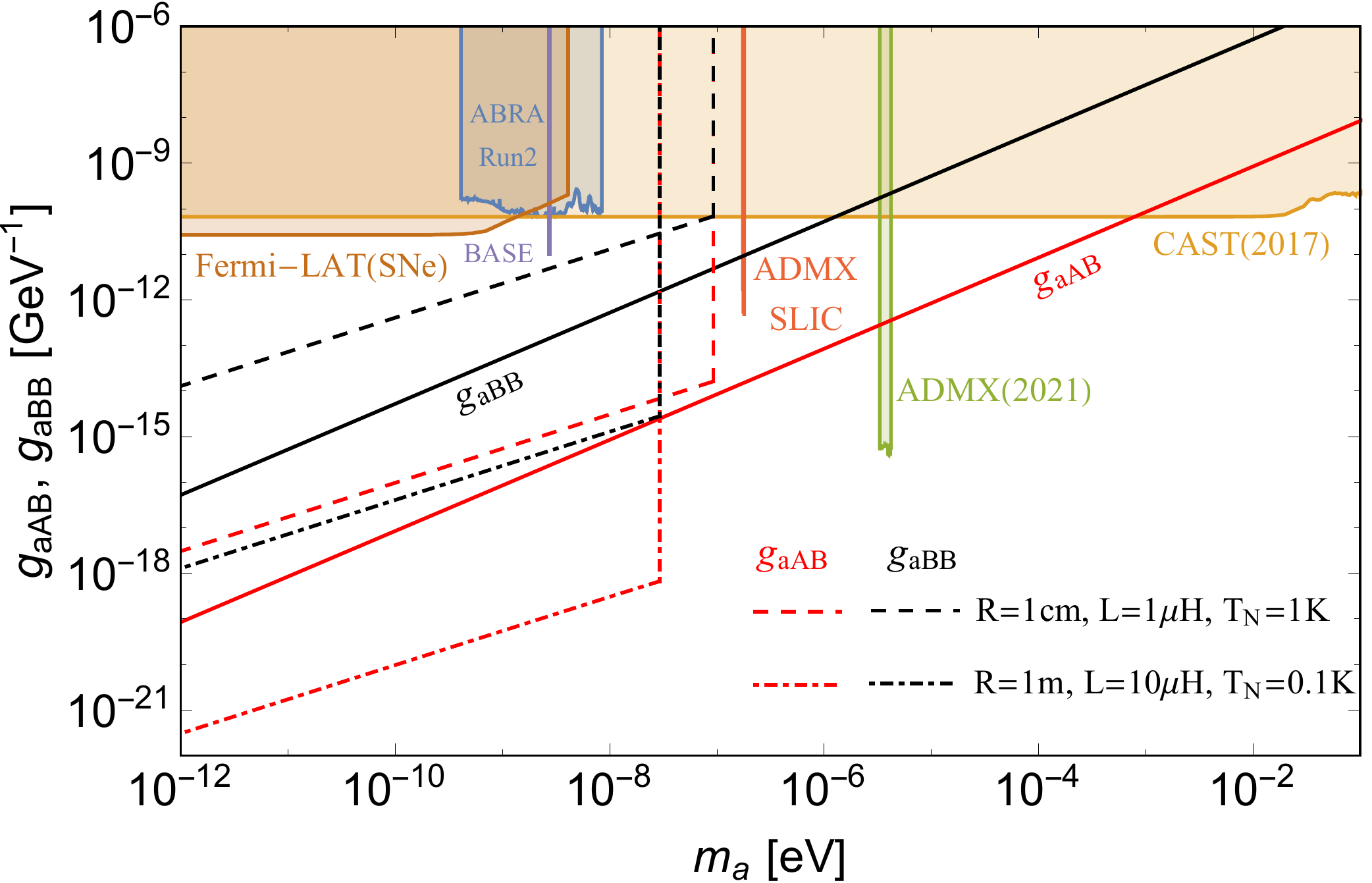}
\caption{The expected sensitivity bounds of $g_{aAB}$ (red lines) and $g_{aBB}$ (black lines). Two setup benchmarks of detector are assumed: $R=1$ cm, $L=1~\mu$H, $T_N=1$ K (dashed) and $R=1$ m, $L=10~\mu$H, $T_N=0.1$ K (dash-dotted). The theoretical predictions of $g_{aAB}$ and $g_{aBB}$ (solid) are also presented~\cite{Sokolov:2022fvs}. Some existing exclusion limits on $g_{a\gamma\gamma}$ are shown for reference, including ABRACADABRA (Run 2)~\cite{Salemi:2021gck}, CAST (2017)~\cite{CAST:2017uph}, ADMX (2021)~\cite{ADMX:2021nhd}, ADMX SLIC~\cite{Crisosto:2019fcj}, BASE~\cite{Devlin:2021fpq}, and Fermi-LAT~\cite{Meyer:2020vzy}.
}
\label{fig:sensitivity}
\end{figure}

The axion-induced electric field in $\phi$ direction $E_{a,\phi}$ is analogous to a vortex electric field produced by the Faraday's electromagnetic induction. We can place a wire loop inside the solenoid to conduct the induction current. The wire loop is then connected in an LC circuit to enhance the signal power. The schematic diagram of experimental setup is shown in Fig.~\ref{fig:setup}. The induction current in a loop of radius $R$ becomes
\begin{eqnarray}
I_a={2\pi R E_{a,\phi}(R)\over R_s}\;,
\end{eqnarray}
where $a_0=\sqrt{2\rho_{\rm DM}}/m_a$ with $\rho_{\rm DM}=0.4~{\rm GeV}~{\rm cm}^{-3}$ being the local DM density, and the resistance is $R_s=L\omega_a/Q_c$ with $Q_c$ as the quality factor of the LC circuit.
The signal power in case I is then given by
\begin{eqnarray}
P_{\rm signal}=\langle I_a^2 R_s\rangle={Q_c\pi^4 g_{aAB}^2 \rho_{\rm DM}B_0^2 R^4 \Big|H_1^+(\omega_a R) J_1(\omega_a R)\Big|^2\over L\omega_a}\;.
\end{eqnarray}
The signal power in case II can be obtained by making a replacement $g_{aAB}B_0\to g_{aBB}E_0$. To measure the signal current, one can adopt either a SQUID magnetometer to pick up the generated magnetic field~\cite{Sikivie:2013laa}, or direct amplifiers to amplify the signal~\cite{Duan:2022nuy}. For the main noise in the signal-to-noise ratio (SNR), we follow the latter method to estimate the thermal noise as
\begin{eqnarray}
P_{\rm noise}=\kappa_B T_N \sqrt{\Delta f\over \Delta t}\;,
\end{eqnarray}
where $\kappa_B$ is the Boltzmann constant, $T_N$ is the noise temperature, $\Delta f=f/Q_c$ is the detector bandwidth and $\Delta t$ is the observation time.   To estimate the sensitivity of $g_{aAB}$ or $g_{aBB}$, we require the SNR to satisfy
\begin{eqnarray}
{\rm SNR}={P_{\rm signal}\over P_{\rm noise}}>3\;.
\end{eqnarray}
The expected sensitivity bounds of $g_{aAB}$ and $g_{aBB}$ are shown in Fig.~\ref{fig:sensitivity}. We assume $Q_c=10^4$~\cite{Sikivie:2013laa}, one week of observation time, and two setup benchmarks for each case with $B_0=14$ T or $E_0=10^3~{\rm kV/m}$. An adjustable capacitance with a minimal value of $50~{\rm pF}$ is set to give a cutoff frequency.

\section{Conclusion}
\label{sec:Con}

The Witten effect implies the electromagnetic interactions between axions and magnetic monopoles. Based on the quantum electromagnetodynamics,
a generic low-energy axion-photon effective field theory was built by introducing two four-potentials ($A^\mu$ and $B^\mu$) to describe a photon. More anomalous axion-photon interactions and couplings ($g_{aAA}$, $g_{aBB}$ and $g_{aAB}$) arise in contrary to the ordinary axion coupling $g_{a\gamma\gamma}aF^{\mu\nu}\tilde{F}_{\mu\nu}$. As a consequence, the conventional axion Maxwell equations are further modified.

In this work we properly solve the new axion-modified Maxwell equations and obtain the axion-induced electromagnetic fields given a static electric or magnetic field.
The induced oscillating magnetic fields are always suppressed compared with the electric fields for the axions with large Compton wavelengths. The dominant couplings $g_{aAB}$ and $g_{aBB}$ can be probed in the presence of external magnetic field and electric field, respectively.

Finally, we propose new strategies to measure the axion-induced electric fields for sub-$\mu$eV axion in haloscope experiments and estimate the sensitivity of $g_{aAB}$ and $g_{aBB}$.

\section*{ACKNOWLEDGMENTS}
T.L. would like to thank Wei Chao and Yu Gao for useful communication.
T.L. is supported by the National Natural Science Foundation of China (Grant No. 11975129, 12035008) and ``the Fundamental Research Funds for the Central Universities'', Nankai University (Grants No. 63196013).

\appendix

\section{The calculation of the anomaly coefficients}
\label{app:coeff}

Ref.~\cite{Zwanziger:1970hk} stated that Zwanziger's QEMD theory introduced a gauge-fixing term to the Lagrangian
\begin{eqnarray}
\mathcal{L}_G={1\over 2n^2} \{[\partial (n\cdot A)]^2+[\partial (n\cdot B)]^2\}\;,
\end{eqnarray}
and thus new equations of motion
\begin{eqnarray}
\partial^2 n\cdot A+\partial^2 n\cdot B=0
\end{eqnarray}
are satisfied.
Then the canonical quantization procedure was performed by identifying the canonical conjugate variables.
As a result, the equal-time canonical commutation relations between the two four-potentials were obtained~\cite{Zwanziger:1970hk}
\begin{eqnarray}
[A^\mu(t,\vec{x}),B^\nu(t,\vec{y})]&=&i\epsilon^{\mu\nu}_{~~\kappa 0} n^\kappa (n\cdot \partial)^{-1}(\vec{x}-\vec{y})\;,\\
~[A^\mu(t,\vec{x}),A^\nu(t,\vec{y})]&=&[B^\mu(t,\vec{x}),B^\nu(t,\vec{y})]=-i(g_0^{~\mu} n^\nu+g_0^{~\nu} n^\mu)(n\cdot \partial)^{-1}(\vec{x}-\vec{y})\;.
\end{eqnarray}
These are the two potentials' non-trivial relations in the QEMD and the right degrees of freedom of photon can be preserved.

Regarding the seeming violation of the Lorentz invariance, Brandt, Neri and Zwanziger formally showed that the observables of the QEMD are Lorentz invariant using the path-integral approach~\cite{Brandt:1977be,Brandt:1978wc}. They claimed that, after all the quantum
corrections are properly accounted for, the dependence on the spatial vector $n_\mu$ in the action
$S$ factorizes into an integer linking number $L_n$ multiplied by the
combination of charges in the quantization condition $q_i g_j - q_j g_i$. This $n$ dependent part is then given by $2\pi N$ with $N$ being an integer. Since $S$ contributes to the generating functional in the form of exponent $e^{iS}$, this
Lorentz-violating part does not play any role in physical processes.

Ref.~\cite{Sokolov:2022fvs} performed the calculation of the anomaly coefficients by following Fujikawa's path integral method~\cite{Fujikawa:1979ay}. We summarize their procedure here.
A high energy QEMD Lagrangian is supposed as
\begin{eqnarray}
\mathcal{L}\supset i\bar{\psi}\gamma^\mu D_\mu \psi + y (\Phi \bar{\psi}_L \psi_R + h.c.)\;,
\label{eq:Lag}
\end{eqnarray}
where the covariant derivative is $D_\mu=\partial_\mu - e q_\psi A_\mu - g_0 g_\psi B_\mu$ with $q_\psi$ ($g_\psi$) being the electric (magnetic) charge of the fermion $\psi$, $\Phi$ is the PQ complex scalar singlet, and $y$ is the Yukawa coupling constant. After the PQ symmetry breaking and a chiral transformation of the fermion in the path integral measure, one has an anomalous term
\begin{eqnarray}
\mathcal{L}_F&=&-{a\over v_{\rm PQ}} \lim_{\Lambda\to \infty, x\to y} {\rm tr}\{\gamma_5 {\rm exp}(\cancel{D}^2/\Lambda^2)\delta^4(x-y) \}\nonumber \\
&=&-{a\over v_{\rm PQ}} \lim_{\Lambda\to \infty, x\to y} \int {d^4 k\over (2\pi)^4} {\rm tr}\{\gamma_5 {\rm exp}[D^2/\Lambda^2-i\gamma_\mu\gamma_\nu (eq_\psi \partial^\mu A^\nu + g_0 g_\psi \partial^\mu B^\nu)/\Lambda^2]e^{ik\cdot (x-y)}\}\;,\nonumber \\
\end{eqnarray}
where a cutoff $\Lambda$ is introduced.
The gauge invariant heat kernel regularization is applied to the path integral.
It turns out that the commutators of the two potentials do not contribute to the derivative square $\cancel{D}^2$ at the same space-time point, i.e., $\cancel{D}^2=D^2-i\gamma_\mu\gamma_\nu (eq_\psi \partial^\mu A^\nu + g_0 g_\psi \partial^\mu B^\nu)$. Then, one can Taylor expand the exponent, and implement the trace, the integral as well as the limit $\Lambda\to \infty$. The anomalous term becomes
\begin{eqnarray}
\mathcal{L}_F&=&{ad(C_\psi)\over 8\pi^2 v_{\rm PQ}} \epsilon_{\mu\nu\rho\sigma} (eq_\psi \partial^\mu A^\nu+g_0g_\psi \partial^\mu B^\nu)(eq_\psi \partial^\rho A^\sigma+g_0g_\psi \partial^\rho B^\sigma)\;.
\end{eqnarray}
Eventually, the anomaly coefficients in Eq.~(\ref{eq:couplings}) can be obtained as
\begin{eqnarray}
E=\sum_\psi q_\psi^2 d(C_\psi)\;,~~M=\sum_\psi g_\psi^2 d(C_\psi)\;,~~D=\sum_\psi q_\psi g_\psi d(C_\psi)\;,
\end{eqnarray}
where $d(C_\psi)$ is the dimension of the color representation of $\psi$.

\bibliographystyle{JHEP}
\bibliography{refs}

\end{document}